\documentclass{aa}
\usepackage{verbatim}
\usepackage{graphicx}
\usepackage{natbib,twoopt} 
\usepackage{array}
\usepackage[breaklinks=true]{hyperref} 
\usepackage{geometry}
\usepackage{multicol}
\usepackage{comment}  
\bibpunct{(}{)}{;}{a}{}{,}     
\makeatletter
  \newcommandtwoopt{\citeads}[3][][]{\href{http://adsabs.harvard.edu/abs/#3}%
    {\def\hyper@linkstart##1##2{}%
     \let\hyper@linkend\@empty\citealp[#1][#2]{#3}}}
  \newcommandtwoopt{\citepads}[3][][]{\href{http://adsabs.harvard.edu/abs/#3}%
    {\def\hyper@linkstart##1##2{}%
     \let\hyper@linkend\@empty\citep[#1][#2]{#3}}}
  \newcommandtwoopt{\citetads}[3][][]{\href{http://adsabs.harvard.edu/abs/#3}%
    {\def\hyper@linkstart##1##2{}%
     \let\hyper@linkend\@empty\citet[#1][#2]{#3}}}
  \newcommandtwoopt{\citeyearads}[3][][]%
    {\href{http://adsabs.harvard.edu/abs/#3}
    {\def\hyper@linkstart##1##2{}%
     \let\hyper@linkend\@empty\citeyear[#1][#2]{#3}}}
\makeatother

\usepackage{graphicx}   
\usepackage{amsmath}    
\usepackage{float}
\usepackage{url}
\usepackage{ulem}
\usepackage{makecell}

\begin{document}
\title{Constraining the formation history of the HAT-P-11 system\\ using atmospheric abundances}

\author{Lena Chatziastros\inst{1}, Bertram Bitsch \inst{1} , Aaron David Schneider \inst{2,3}
}

\institute{Max-Planck-Institut für Astronomie, Königstuhl 17, 69117 Heidelberg, Germany
            \and
            Niels Bohr Institutet, Københavns Universitet, Blegdamsvej 17, 2100 København, Denmark
            \and
            Institute of Astronomy, KU Leuven, Celestijnenlaan 200D, 3001, Leuven, Belgium
             }

\date{Received ; accepted }
 
\abstract
{The chemical fingerprint of a planet can reveal information about its formation history regarding when and where the planet formed. In particular, the water content of a planet can help to constrain its formation pathway: If the planet formed in the outer regions of the disk and migrated inward, it would be water-rich due to the accretion of water-ice-rich solids. Conversely, formation in the inner disk region, where water-ice is not available, would result in a smaller atmospheric water content due to the limited accretion of water vapor. However, this process becomes complex with the presence of gap-opening giant planets. A gas giant exerts a pressure bump exterior to its orbit, preventing further influx of pebbles into the inner system, resulting in a water-poor environment and eventually leading to water-poor inner planets. These different formation scenarios can help to constrain the formation of the HAT-P-11 system, which contains an inner sub-Neptune with a mass of 23.4 $\mathrm{M_{\oplus}}$ and substellar water abundances ($X_\mathrm{H_2O} \approx 0.11$), as well as an outer giant planet orbiting exterior to the water-ice line. Our planet formation model encompasses planetary growth through pebble and gas accretion, along with a pebble drift and evaporation module that enables us to track the chemical composition of the disk and the planet over time. We find that the presence of the gas giant is necessary to block water-ice-rich material, resulting in a substellar water content for the inner sub-Neptune, HAT-P-11b. On the other hand, if the giant planet forms too early, not enough solid material can enter the inner disk regions, preventing efficient growth of the inner planet. This highlights the importance of the timing of giant planet formation in explaining the inner system structure, including the formation of Jupiter in our Solar System. Furthermore, our simulations predict a roughly stellar C/O ratio with superstellar C/H and O/H ratios for HAT-P-11b, providing constraints for future observations of this system, which are essential for gaining a more detailed understanding of its formation.}

\keywords{planet formation --
          planet-disk interactions --
          planetary atmospheres}

\authorrunning{L. Chatziastros et al.}
\titlerunning{Constraining the formation history of the HAT-P-11 system using
atmospheric abundances}
\maketitle 

\section{Introduction} \label{introduction}
In recent years, a great number of exoplanets have been discovered, and these planets exhibit great variation in their size, mass, chemical composition, and distance to their central star. When constraining the potential planet-formation histories of known exoplanets, the primary factors considered are their size, mass, and distance to their central star. However, the chemical fingerprint of a planet can provide essential information about its formation history \citep{oberg2011effects, mordasini2016imprint, madhusudhan2017atmospheric, cridland2019connecting, schneider2021drifting, bitsch2022drifting, molliere2022interpreting}. The planet's composition can offer insights into when the planet formed, how far it migrated in the protoplanetary disk, and the impact of other planets in the system.\\
\cite{bitsch2021dry} introduced the idea that early-forming protoplanetary gas giants significantly influence the chemical composition of the disk within their orbit by blocking pebbles beyond their orbit (e.g., \citealt{paardekooper2006dust, lambrechts2014separating, morbidelli2016fossilized, ataiee2018much, bitsch2018pebble, weber2018characterizing}). Depending on their position in the disk, gas giants can obstruct the inward drift of material in cold regions of the disk, such as at large radii, and prevent its drift to hotter regions and subsequent evaporation, resulting in volatile-poor inner disk regions. In turn, the composition of the disk material affects the potential composition of a planet forming within it.\\
As presented in \cite{bitsch2022drifting}, the chemical abundances of hot Jupiters WASP-77A b \citep{line2021solar} and $\tau$ Boötis b \citep{pelletier2021water} can already be reproduced within simulations that consider the effects of pebble drift and evaporation at ice lines, while it is assumed that these planets are solitary. However, with a second massive planet, the development of the composition of the disk changes because of the blocking of inwardly drifting material.\\
In this study, we focus on investigating how the water content of a planetary envelope is affected by the presence of a gas giant in the protoplanetary disk, specifically in the planetary system HAT-P-11. The HAT-P-11 system is currently the only known planetary system with an outer gas giant and an inner sub-Neptune for which there are constraints on the atmospheric water content of the inner planet \citep{welbanks2019mass}. The inner sub-Neptune, HAT-P-11b, is located at a distance of 0.053 AU to the central star, while the gas giant, HAT-P-11c, is at a distance of 4.13 AU to the central star (see The Extrasolar Planets Encyclopaedia\footnote{The Extrasolar Planets Encyclopaedia: \\ \url{http://www.exoplanet.eu}; accessed on 2022-03-28}). Therefore, we investigate different formation scenarios in regard to the timing and location of the formation of the  inner sub-Neptune and how these factors influence its water content, as similarly suggested in \citeauthor{bitsch2021dry}\\ (\citeyear{bitsch2021dry}). According to observational data, the inner planet HAT-P-11b appears to have a substellar water content compared to the abundances of the  central star (see \citealt{welbanks2019mass} for details). We compare the simulation results provided chemcomp (code described in \citealt{schneider2021drifting}) to observational data, providing strong constraints on the formation history of the  system. \\
\\
This paper is structured as follows: we present the numerical
methods we use in Sect. \ref{methods}, and the results in Sects. \ref{scen1} and \ref{scen2}. We then discuss the obtained results in Sect. \ref{discussion}, before concluding in Sect. \ref{conclusion}.

\section{Methods} \label{methods}
The simulations for this study are made with \texttt{chemcomp} (see \citealt{schneider2021drifting} for a detailed description of the code). 
This code includes viscous disk evolution (e.g., \citealt{lynden1974evolution}) combined with a recipe for dust growth and drift (following \citealt{birnstiel2012simple}), as well as pebble evaporation and recondensation at ice lines (e.g., \citealt{piso2015c, ros2019effect, schneider2021drifting}). In addition, planet formation is simulated by pebble accretion \citep{johansen2017forming} and gas accretion (e.g., \citealt{ndugu2021probing}), as well as migration \citep{paardekooper2011torque} and gap opening in the disk \citep{crida2006width}. We assume that, during pebble accretion, 10\% of the accreted solids contribute to the primordial envelope of the planet, creating a heavy-element atmosphere, as expected from more detailed simulations of planetary envelopes (e.g., \citealt{brouwers2021planets}). A planet forming exterior to the water-ice line can therefore contain water vapor in the planetary atmosphere due to water-ice accretion. A detailed description of chemcomp is provided in \cite{schneider2021drifting}.\\
Some adjustments
were made to adapt the chemcomp code to the question under investigation here. In particular, we modified the chemical composition and chemistry model of the disk. Additionally, we implement the impact of a gas giant on the disk by allowing it to open a gap in the disk. These modifications are explained in more detail in  Sects. \ref{protoplanetary disk} and \ref{gas giant}.

\subsection{Protoplanetary disk} \label{protoplanetary disk}
For the setup of the protoplanetary disk, we followed \cite{schneider2021drifting}. The following parameters were used: the mass of the disk is set to $M_0$ = 0.128 $\mathrm{M_{\odot}}$, the disk radius is $R_0$ = 137 AU, and the metallicity amounts to $\epsilon_d$ = 2\%.  \cite{savvidou2023make} (in review) state that disks with $M_0$ = 0.1 $\mathrm{M_{\odot}}$ are in favor of producing gas giants, as one is present in the HAT-P-11 system. Furthermore, the stellar mass and luminosity were adjusted. Therefore, the stellar mass of HAT-P-11 $M_\mathrm{stellar}$ = 0.809 $\mathrm{M_{\odot}}$ as indicated in \cite{hinkel2014stellar} and the stellar luminosity of $L_\mathrm{stellar}$ = 1 $\mathrm{L_{\odot}}$ was chosen, where the luminosity is fixed over time; 1 $\mathrm{L_{\odot}}$
is the temporal mean luminosity of a star with this particular mass (see \citealt{baraffe2015new}). To set the initial distribution of the chemical composition in the protoplanetary disk, we use the stellar abundances of HAT-P-11 as obtained from the Hypatia Catalog \citep{hinkel2014stellar}. We show these values in Table \ref{tab:stellar abundances}. No data are available for [K/H] and [S/H], even though they are included in the original chemical model used in chemcomp \citep{schneider2021drifting}. Therefore, [K/H] was neglected, and [S/H] was set to the value of [Si/H], as the abundances of these elements are correlated \citep{chen2002sulphur}. \\

\begin{table*}[h]
   \centering
    \begin{tabular*}{\textwidth}{@{\extracolsep{\fill}}c|c|c|c|c|c|c|c|c|c|c|c|c}
    \hline
    \hline
    \rule{0pt}{9.5pt}\small{Element} & 
    $\mathrm{[O/H]}$  &
    $\mathrm{[C/H]}$ &
    $\mathrm{[N/H]}$ &
    $\mathrm{[Mg/H]}$ & 
    $\mathrm{[Si/H]}$ &     
    $\mathrm{[Fe/H]}$ & 
    $\mathrm{[S/H]}$ &
    $\mathrm{[Al/H]}$ & 
    $\mathrm{[Na/H]}$ & 
    $\mathrm{[K/H]}$ & 
    $\mathrm{[Ti/H]}$ & 
    $\mathrm{[V/H]}$ \\
    \hline
    \rule{0pt}{9.5pt}\small{Abund.} & 
    0.04 & 0.14 & 0.32& 0.28 & 0.32 &
    0.41 & 0.32 & 0.34 & 0.43 & - & 0.36 & 0.48\\
    \hline
    \hline
    \multicolumn{13}{c}{}
    \end{tabular*}
    \caption{Observational data of elemental abundances of HAT-P-11 normalized to solar values were obtained from the Hypatia Catalog Database \citep{hinkel2014stellar}. No constraints are available for [K/H] and [S/H]. Therefore, [K/H] was neglected, and [S/H] is assumed to scale as [Si/H], as motivated by observations \citep{chen2002sulphur}.}
    \label{tab:stellar abundances}
\end{table*}
\noindent For the development of the element distribution in the protoplanetary disk over time, we use the chemistry model described by \cite{bitsch2020influence}, \cite{schneider2021drifting}, and \cite{schneider2021drifting2}. In this model, the elements are distributed in different proportions to the molecules to be formed. Because the disk temperature is dependent on radial distance to the star, molecules of a certain species are either in gaseous form or in solid form, depending on their position in the disk and their specific evaporation temperature. The position in the disk where the evaporation temperature of a certain species is equal to the disk temperature is called the specific evaporation line of that species. Due to drifting pebbles over these evaporation lines, the initial chemical composition of the disk changes with time.\\
Here, two distinct chemistry models exist, which differ mainly in the proportion of carbon grains, that is, 20\% versus 60\% carbon grains \citep{bergin2015tracing, altwegg2020evidence}. Using the stellar abundances of HAT-P-11 (Table \ref{tab:stellar abundances}) and the corresponding chemical partitioning model would not allow the formation of water if only 20\% of the carbon was in refractory grains. This is because, with a lower number of carbon grains, more carbon molecules are bound in $\mathrm{CO}$ and $\mathrm{CO_2}$, taking away the oxygen and preventing the formation of water. Consequently, the chemistry model with 60\% carbon grains is used (see Table \ref{tab:chem model} for details). \\
We define the water content of the star as the ratio of the number of molecules of water and hydrogen:
\begin{align}
      X_{\mathrm{H_2O, \hspace{2px} stellar}} = \frac{N_\mathrm{{H_2O}}}{N_\mathrm{H}}
      = X_\mathrm{H_2O, \hspace{2px} disk, \hspace{2px} t=0} \hspace{4px} \text{,}
\end{align}
where $X_\mathrm{H_2O, \hspace{2px} disk, \hspace{2px} t=0}$ corresponds to the water content in the disk at time zero and $N_\mathrm{i}$ is the number of molecules for a certain species. As pebbles drift inward and release their volatiles into the disk, the water content of the disk changes over time (e.g., \citealt{piso2015c, schneider2021drifting, bitsch2021influence}). The water content of the gas that is then accreted by the planet is determined by the water content of the disk, ultimately setting the water abundance of the atmosphere of the planet. We define the water content in general as
\begin{align}
    X_\mathrm{H_2O} \left(\mathrm{r,t} \right) = \frac{X_\mathrm{H_2O, \hspace{2px} disk} \left(\mathrm{r,t} \right) }{X_\mathrm{H_2O, \hspace{2px} stellar}} \hspace{4px} \text{.}\label{eq:xh2o}
\end{align}
\cite{welbanks2019mass} suggest that the water vapor content in the envelope  of HAT-P-11b is substellar with a value of
\begin{align}
    X_\mathrm{H_2O} = 0.11^{+0.61}_{-0.08} \hspace{4px} \text{.}
\end{align}
Our viscous disk evolution model follows the $\alpha$-viscosity prescription of \cite{shakura1973black}. The viscosity $\nu$ for the disk is then given as
\begin{align}
    \nu =  \alpha \hspace{2px} \frac{c_s^2}{\Omega} \hspace{4px} ,
\end{align}
where $\alpha$ is a dimensionless parameter in the range of $10^{-4}$ to $10^{-2}$, $c_s$ describes the speed of sound, and $\Omega(r)$ is the Keplerian angular frequency, defined as
\begin{align}
    \Omega(r) = \sqrt{\frac{G M_{\star}}{r^3}} \hspace{4px} ,
\end{align}
with the gravitational constant $G$, the mass of the central star $M_{\star}$, and the disk radius $r$.

\subsection{Planet formation of the sub-Neptune}
After the planet has reached pebble isolation mass (e.g., \citealt{lambrechts2014separating, ataiee2018much, bitsch2018pebble}), it can slowly begin to contract its envelope. The gas accretion is determined by the envelope contraction rate of the planet, which in turn depends on the envelope opacity $k_\mathrm{env}$ (e.g., \citealt{ikoma2000formation, mordasini2014grain, brouwers2021planets}), with typical values on the order of $10^{-3}$ to $10^{-2}$  cm$^2$/g. However, such values result in very rapid envelope contraction rates leading to transitions into gas giants (e.g., \citealt{ikoma2000formation, bitsch2021influence}). In this study, we aim to investigate the formation of a Neptune-sized planet, which requires a much slower transition into runaway gas accretion. Therefore, we use an envelope opacity of $k_\mathrm{env}$ = 100 cm$^2$/g for scenario 1 and $k_\mathrm{env}$ = 10 cm$^2$/g for scenario 2 (details in Sects. \ref{scen1} and \ref{scen2}) to prevent the transition into gas giants. It is important to note that this is just a proxy parameter, and in reality, more complex mechanisms 
could prevent atmospheric contraction, such as planetesimal bombardment (e.g., \citealt{alibert2018formation}) or recycling flows (e.g., \citealt{cimerman2017hydrodynamics, lambrechts2017reduced}).

\subsection{Gas giant} \label{gas giant}
In this study, we are interested in the influence a gas giant can have on the chemical composition of the disk, and consequently on the planet evolving within it. The key effect to be simulated is the gravitational impact of the gas giant on the disk. The gap in the disk is primarily influenced by the  mass of the gas giant.\\
For a planet to create a gap in the disk, the following condition must hold, following \cite{crida2006width, armitage2010astrophysics}:
\begin{equation}
    P = \frac{3}{4}\frac{H_{g}}{R_H} + \frac{50}{q \mathcal{R}} \leq 1 \hspace{4px} .
    \label{eq:bedingung lücke}
\end{equation}
Here, $H_{g}$ represents the scale height of the disk, $R_H = a_P \left(\frac{\bold{M_P}}{3M_{\star}}\right)^{1/3}$ stands for the planetary Hill radius, and $\mathcal{R} = a_P^2\Omega / \nu$ is the Reynolds number. The planet's orbital position in the disk is denoted $a_P$, the mass of the planet is described by $M_P$, and $q = \frac{M_P}{M_{\star}}$ is the mass ratio of the planet to the central star.
The depth of the gap caused by the gravitational influence of the planet can be calculated as follows:
\begin{equation}
   f(P) = \begin{cases}
                \frac{P-0.541}{4} & \text{for $P$ $<$ 2.4646} \\\
                1.0 - \exp \left(-\frac{P^{3/4}}{3}\right) & \text{else,}\\
           \end{cases}
    \label{eq:grav gap}
\end{equation}
where the depth of the gap corresponds to the surface density ratio $\Sigma / \Sigma_0$ of the disturbed region to the undisturbed disk. A gap is considered fully open when $\Sigma / \Sigma_0 \leq$ 0.1.\\
To mimic the effect of a gas giant on the disk without simulating the entire planetary evolution, only its gravitational effect is computed over time. Calculating the gravitational influence of the developing gas giant requires tracking the evolution of its mass. To achieve this, we fully
simulated a gas giant, neglecting migration. We then fitted a growth function to describe the mass growth of the  gas giant over time. This fitted mass growth function is then used to calculate the influences of the gas giant on the surface density of the disk. Thus, the gap depth (Eq. \ref{eq:grav gap}) is calculated at each time step at the theoretical position of the  gas giant  in the disk, which we assume to be fixed over time for simplicity. This approach yields satisfactory results when compared to the fully simulated gas giant (for details see Appendix \ref{appendix: gas giant}). The gas giant must begin its growth early to become massive enough, and so growth begins at $t_0 = 0.0$ Myr throughout this study.

\subsection{Formation scenarios}
We examine two different formation scenarios for the HAT-P-11 system, focusing on the starting positions of the inner sub-Neptune. In the first scenario, a solitary planet evolves beyond the water-ice line and migrates inward during its formation process, eventually crossing the water-ice line. In this scenario, the planet can accrete water-rich pebbles beyond the water-ice line and water vapor within the water-ice line. The results for this scenario are presented in Sect. \ref{scen1}.\\
In scenario 2, the sub-Neptune evolves within the water-ice line. Additionally, a gas giant develops in the protoplanetary disk, remaining exterior to the water-ice line at all times, and planet migration is neglected. In this scenario, the growing outer gas giant can create a pressure bump and block the inwardly drifting water-ice-rich pebbles, ultimately resulting in a water-poor inner disk. The  atmospheric water content of the  inner sub-Neptune is then solely determined by vapor accretion. Results are shown in Sect. \ref{scen2}.\\
If the giant planet forms within the water-ice line, it has no impact on the water content of the  disk, as its pressure bump does not block inwardly drifting water-ice-rich pebbles beyond their evaporation line, and their evaporation is therefore not prevented (see Appendix \ref{appendix: scen3}).

\section{Scenario 1: Planet formation without gas giant} \label{scen1}
\subsection{Water vapor content in the protoplanetary disk}

\begin{figure*}[h]
    \centering
    \includegraphics[width=\textwidth]{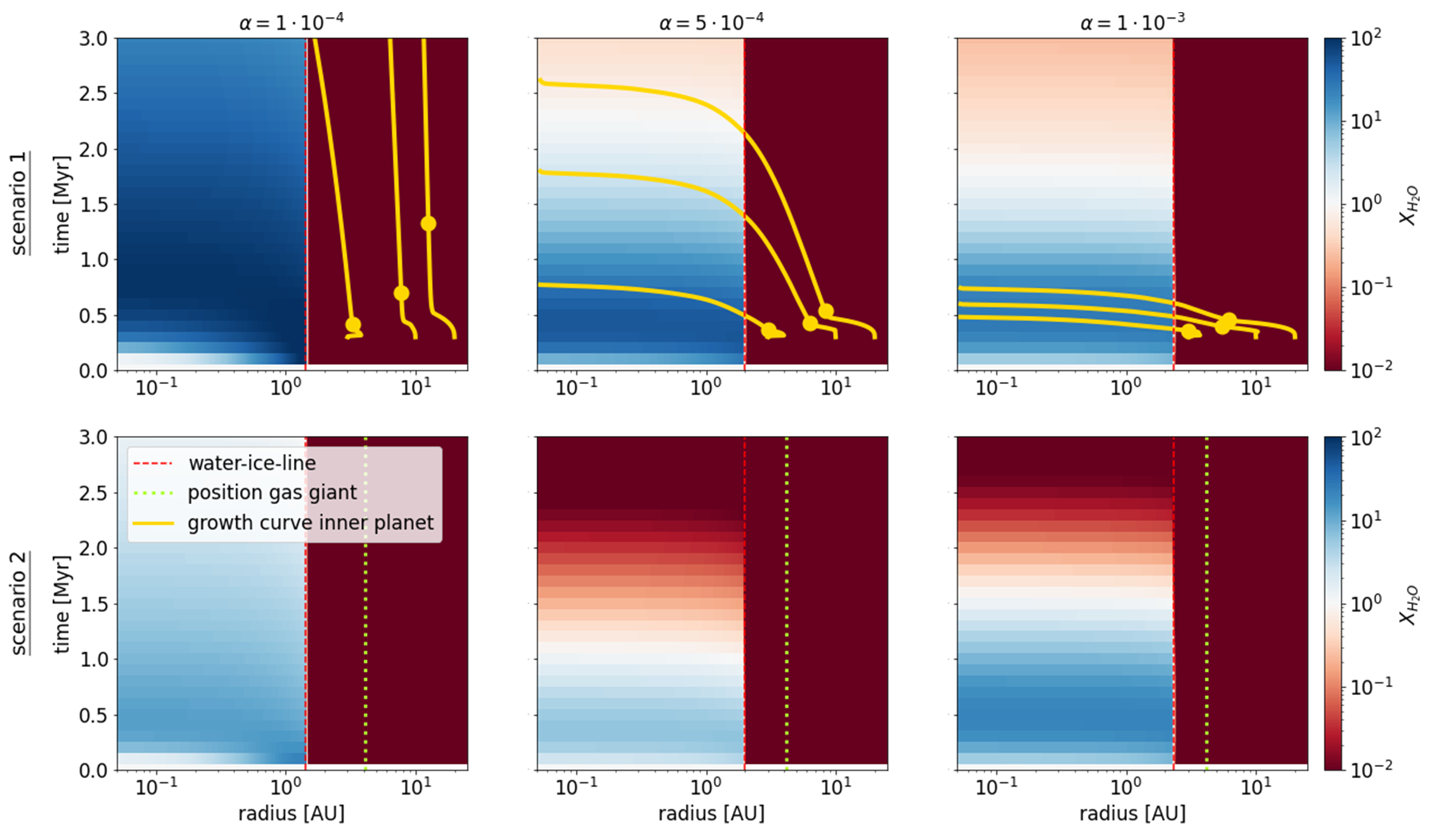}
    \caption{Water content in the gas phase within the water-ice line as a function of disk radius and time, normalized to stellar composition. Results for different $\alpha$-viscosities for scenario 1 (top) and scenario 2 (bottom) are displayed. Blue regions indicate superstellar water content, while red regions indicate substellar water content. Beyond the water-ice line (marked with a red vertical line), water exists only as ice, resulting in zero water vapor abundance exterior to the ice line. For scenario 1, we additionally show the growth tracks of the planets inserted at 0.3 Myr, where the yellow dot indicates the time when the planet reaches pebble isolation mass and enters the gas-accretion phase. For scenario 2 (bottom), the position of the gas giant is marked with a green vertical line.}
    \label{fig:water content disk + growthtracks scen 1}
\end{figure*}

To estimate the potential water content for a planet forming within this disk, we first investigated the water vapor content of the protoplanetary disk. Using the methods described above, we conducted simulations of the protoplanetary disk without a planet present in order to observe the evolution of the water vapor content in the disk. The results are depicted in Fig. \ref{fig:water content disk + growthtracks scen 1} (top). The water content in the disk is displayed for three different $\alpha$-viscosities as a function of disk radius and time. Here, the water content is normalized to the composition of the central star (see Eq. \ref{eq:xh2o}) to allow direct comparisons with the results of \cite{welbanks2019mass}.\\
In Fig. \ref{fig:water content disk + growthtracks scen 1}, the water vapor content of the disk is shown as a function of disk radius and time. 
It can be observed that for disks with $\alpha = 1 \cdot 10^{-3}$ and $\alpha = 5 \cdot 10^{-4}$, the water vapor content in the disk is superstellar (blue) at first and becomes substellar (red) at later times. For $\alpha = 1 \cdot 10^{-4}$, a substellar water vapor content is not reached within the lifetime of the disk.\\
The evolution of water content proceeds at different rates for disks with varying viscosities. The change in water vapor content is primarily caused by the evaporation of ice-bearing pebbles. At the beginning of the disk's lifetime, these pebbles drift across the evaporation line and gradually evaporate over time. During this process, the gas in the inner disk is initially enriched with water vapor. Once the supply of water-ice-bearing pebbles is depleted, no further water can evaporate. As the gas gets accreted toward the central star over time, the water-rich gas is slowly depleted in regions within the water-ice line. Additionally, carbon-bearing volatiles (e.g., $\mathrm{CO_2}$, $\mathrm{CH_4}$ and CO) gradually drift inward, altering the C/O ratio of the disk. Pebbles become larger at lower viscosities and therefore drift inward more quickly \citep{birnstiel2012simple}, while gas transport is slower for low-$\alpha$ disk. Consequently, in low-viscosity disks, the entire supply of ice-rich pebbles evaporates more rapidly, but gas transport is also much slower, which hinders the efficient reduction of high water vapor content. Therefore, the disk with $\alpha = 1 \cdot 10^{-4}$ does not have sufficient time to remove the water-rich gas during its lifetime of 3 Myr but would only do so after approximately 10 Myr.

\begin{figure*}[h]
    \centering
    \includegraphics[width=\textwidth]{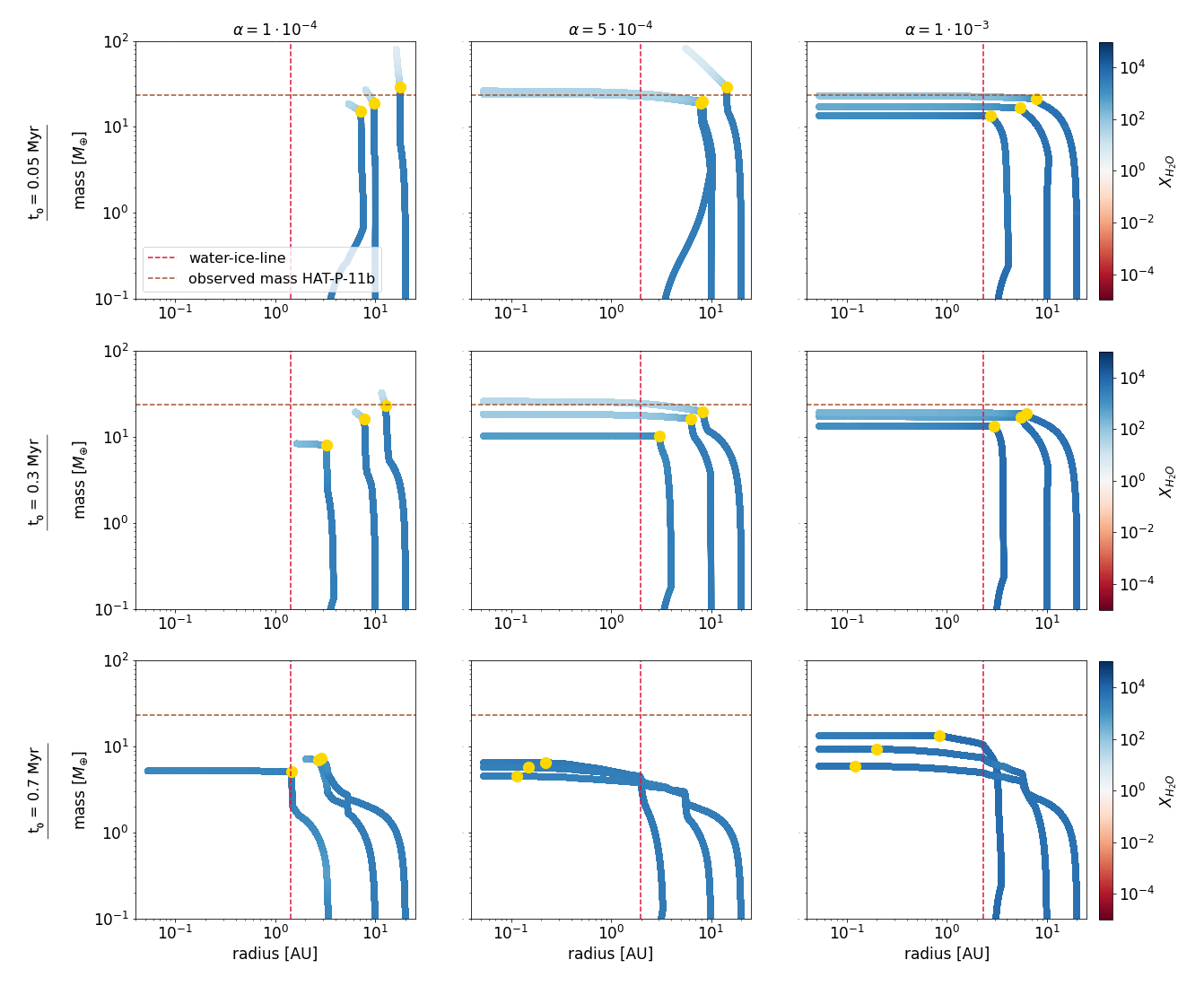}
    \caption{Growth track of the inner planet for different insertion times (0.05 Myr, 0.3 Myr, 0.7 Myr, top to bottom), initial positions (3 AU, 10 AU, 20 AU), and $\alpha$-viscosities ($\alpha =  1 \cdot 10^{-4}$,  $5 \cdot 10^{-4}$, $1 \cdot 10^{-3}$, left to right) as a function of the  position of the planet in the disk and its total mass. The color code represents the water content of the planetary envelope normalized to stellar composition. The yellow dot indicates the onset of gas accretion, the red vertical line marks the water-ice line, and the orange horizontal line indicates observed mass of  HAT-P-11b, namely 23.4 $\mathrm{M_{\oplus}}$. Each simulation results in superstellar water content for the planetary envelope, contrary to the strongly substellar water content of HAT-P-11b.}
    \label{fig:water content planet scen 1}
\end{figure*}

\begin{figure*}[h]
    \centering
    \includegraphics[width=1\textwidth]{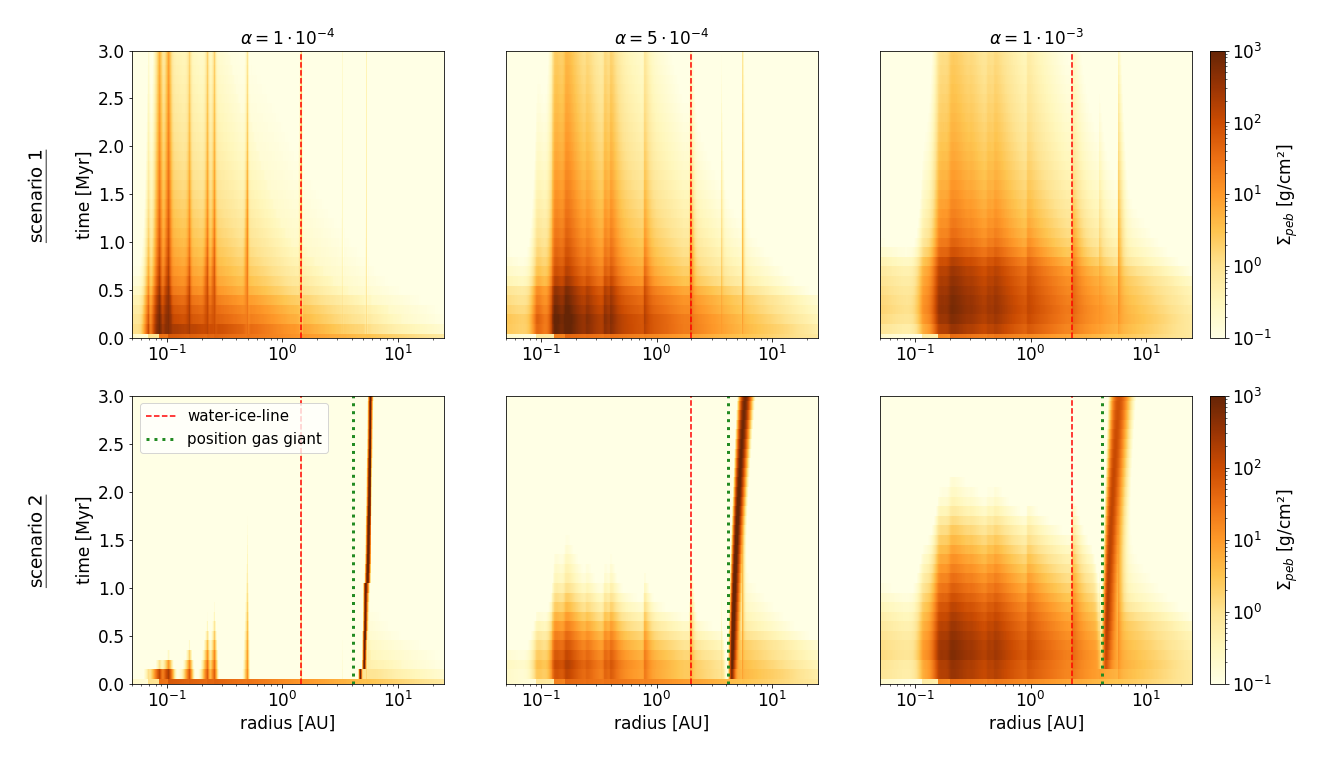}
    \caption{Pebble surface density for scenario 1 (top) and scenario 2 (bottom) for different $\alpha$-viscosities, as a function of disk radius and time. In each case, the red vertical line denotes the water-ice line, and the green line marks the position of the gas giant, if present. In the scenario with a gas giant, the accumulation of blocked pebbles exterior to the gas giant's position is visible. Further accumulations within the water-ice line result from the recondensation of outwardly diffusing gas at different evaporation lines.}
    \label{fig:dust surface density}
\end{figure*}

\subsection{Water content in the planetary envelope}
In the first planet formation scenario, a single planet forms exterior to the water-ice line. As the planet grows and migrates inward over time, it eventually crosses the water-ice line. For simplicity, we assume that the inner planet formed before the gas giant, allowing us to solely simulate the growth of the inner planet. We test different initial positions (3 AU, 10 AU, 20 AU) and initial times ($t_0=$ 0.05 Myr, 0.3 Myr, 0.7 Myr). Initiating the planetary growth at later times ($t_0 >$ 0.7 Myr) results in inefficient planetary growth because most pebbles have already drifted inward, leaving very little material to be accreted by the planet.\\
In Fig. \ref{fig:water content disk + growthtracks scen 1}, the planetary growth tracks for planets inserted at $t_0$ = 0.3 Myr are shown in orange curves, with the dot indicating the moment when the planet reaches pebble isolation mass, signifying the onset of gas accretion. We clearly observe that the planet migrates inward more rapidly with increasing viscosity. At higher viscosities, a planet must reach higher mass to open a deep gap and transition into the slower type-II migration (e.g., \citealt{crida2006width, kanagawa2018radial, bergez2020influence}). In contrast, at low viscosity, even planets with 20 Earth masses can start opening gaps, reducing their inward migration speed compared to planets of similar mass in high-viscosity disks, which still migrate in type-I. In individual simulations, the growing planet may initially migrate outward due to the heating torque before the Lindblad torque becomes dominant and causes inward migration \citep{benitez2015planet, masset2017coorbital, guilera2019thermal, baumann2020influence}. The heating torque is most effective during periods of rapid pebble accretion (at low viscosity) because the planet is very hot during this process and therefore has a high luminosity \citep{chrenko2017eccentricity}.\\
\enlargethispage{15px}
In general, the simulation terminates as soon as the planet reaches the inner edge of the disk, defined at a distance of 0.053 AU from the central star, corresponding to the observed position of HAT-P-11b. However, the planet does not reach the observed final position in simulations with $\alpha = 1 \cdot 10^{-4}$. The final orbital position of the planet could be achieved if the disk were to last longer or if scattering events were to take place with the outer gas giant.\\ 
Moreover, there is a tendency for late-forming planets to be longer in the pebble-accretion phase. This is because, at later times, some pebbles have already drifted inward, reducing the available material in the disk for planet accretion. Consequently, pebble-accretion rates decrease, and planets take longer to reach pebble isolation mass.\\ 
Additionally, Fig. \ref{fig:water content planet scen 1} displays the water content of the planet's envelope for all 27 simulations of this scenario as a function of the position of the planet in the disk and its total mass. The color coding represents the normalized water content of the planetary envelope. In this scenario, for most simulations, the majority of the planetary envelope mass is acquired during the pebble-accretion phase. As the planet forms beyond the water-ice line, it accretes ice-bearing pebbles, enriching the envelope with water before gas accretion commences. This results in a relatively water-rich envelope, as 10\% of the accreted solid material through pebble accretion contributes to the protoplanetary envelope, while this material is by construction very water-rich. As gas accretion progresses, the already high water content cannot be reduced to substellar values.\\
Also, the final mass of the  planet depends on the time of formation. An early-forming planet becomes more massive than a late-forming planet because of pebble drift, regardless of the disk viscosity. Especially in low $\alpha$ disks, pebbles rapidly drift inward and provide material early in the inner regions of the disk. However, this also implies that pebbles in the far outer regions drift inward quickly, making it challenging for massive planets to form there. This becomes clear when examining the surface density of pebbles in Fig. \ref{fig:dust surface density} (scenario 1: top row). The color code indicates the pebble surface density in the disk, varying with disk radius and time. We clearly see that both beyond and within the water-ice line, the pebble surface density is significantly reduced early on, especially at $\alpha = 1 \cdot 10^{-4}$. The accumulation of pebbles observed here at certain radii is a result of outward-diffusing gas, which recondenses in the process (see also \citealt{schneider2021drifting}). For planet formation times $t_0 >$ 0.7 Myr, it is highly unlikely that the planet reaches several Earth masses. Further simulations with even later planet-formation times result in masses smaller than 3 Earth masses and are therefore not of interest for the formation of HAT-P-11b. Nevertheless, in scenario 1, the observed mass of HAT-P-11b, which is of 23.4 $\mathrm{M_\oplus}$, can be reproduced in some simulations.\\
This scenario reveals that the pebble-drift velocity limits potential planet formation times. However, the water content of the planetary atmosphere obtained in these simulations is too high compared to observations. Consequently, it seems highly improbable that HAT-P-11b formed in the outer disk and then subsequently migrated inward.

\section{Scenario 2: Planet formation with gas giant exterior to the water evaporation line} \label{scen2}
To investigate the influence of a gas giant on the water content in the disk, a gas giant was positioned exterior to the water-ice line for the second scenario. The implementation of the gas giant in the code is described in Sect. \ref{methods}. Here, we chose the observed position of HAT-P-11c at 4.13 AU for all viscosities. In reality, planets migrate, but what is crucial here is that the planet remains exterior to the water-ice line to block the inward drift of pebbles. For the sake of simplicity, we assume that the planet always remains at 4.13 AU, aligning with its present-day observation.

\subsection{Water vapor content in the protoplanetary disk}
Figure \ref{fig:water content disk + growthtracks scen 1} (bottom) illustrates the water content of the disk with the gas giant positioned beyond the water-ice line for different viscosities. In comparison to simulations without a gas giant (Fig. \ref{fig:water content disk + growthtracks scen 1}, top), a distinct variation in water content can be seen at later stages. The blocking of drifting pebbles exterior to the water-ice line is also clearly depicted in the surface density of the  disk in Fig. \ref{fig:dust surface density} (bottom). The gas giant attains pebble isolation mass after between 0.2 and 0.5 Myr, depending on the $\alpha$-viscosity (see Appendix \ref{appendix: gas giant} for details), which determines when pebble blocking becomes efficient. At later stages, the surface density of pebbles in the inner disk significantly decreases. This effect is more pronounced at lower viscosities due to the earlier formation of the gap by the growing giant and the faster drift of pebbles caused by larger grains at lower viscosity (e.g., \citealt{birnstiel2012simple}). The gap is formed earlier at lower viscosity because the pebble isolation mass is lower \citep{ataiee2018much, bitsch2018pebble}, and accretion rates are higher, and therefore the drifting pebbles are blocked earlier compared to the case with higher viscosity. Consequently, the influence of the gas giant becomes evident earlier in low-viscosity disks. Although a noticeable reduction in the water content is observed for $\alpha = 1 \cdot 10^{-4}$, a substellar water content is not achieved. This is due to the inefficiency of gas diffusion and therefore of removing the water-rich gas throughout the  lifespan of the disk.\\
In contrast, for $\alpha = 1 \cdot 10^{-3}$ and $\alpha = 5 \cdot 10^{-4}$, substantially substellar water contents are reached in the disk even earlier than in scenario 1. This is attributed to the prevention of inward drift and evaporation of water-ice-rich pebbles. Additionally, at these higher viscosities, water vapor is efficiently transported away. Substellar water contents are reached earlier in a disk with $\alpha = 5 \cdot 10^{-4}$ than in one with $\alpha = 1 \cdot 10^{-3}$, because of earlier gap opening by the gas giant in low-viscosity disks (see Appendix \ref{appendix: gas giant}).

\subsection{Water content of the inner planet}
\begin{figure}[t]
    \centering
    \includegraphics[width=0.5\textwidth]{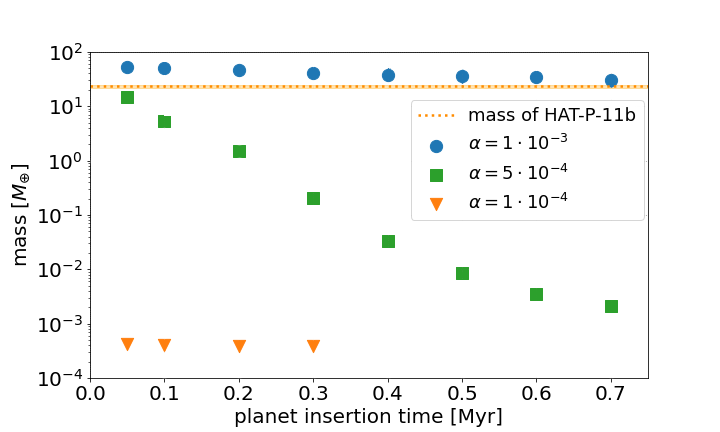}
    \caption{Final masses of the inner planet are shown for different planet insertion times and different $\alpha$-viscosities of the disk. The orange horizontal line represents the observed mass of  HAT-P-11b (23.4 $\mathrm{
    M_{\oplus}}$) within its range of uncertainty.}
    \label{fig:planet mass scen 3}
\end{figure}
\begin{figure}[t] 
    \centering
    \includegraphics[width=0.5\textwidth]{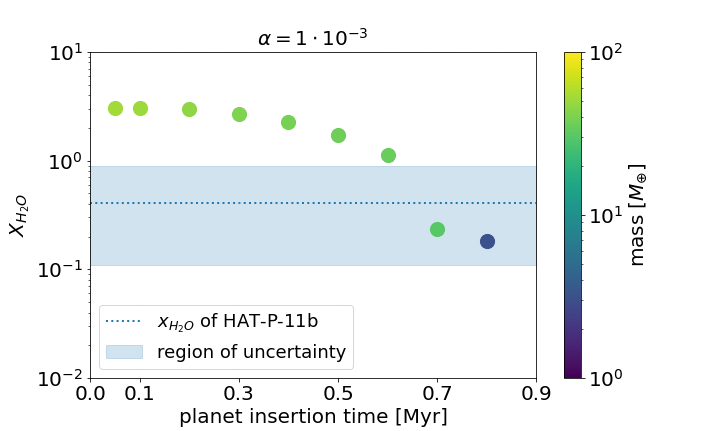}
    \caption{Final water content of the envelope of the  inner planet at different insertion times in a disk with $\alpha = 1 \cdot 10^{-3}$ , normalized to stellar abundances. The color code indicates the final planetary mass and the blue line represents the observed water content of HAT-P-11b within its region of uncertainty.}
    \label{fig:water content planet scen 3}
\end{figure}
\begin{figure}[t]
    \centering
    \includegraphics[width=0.5\textwidth]{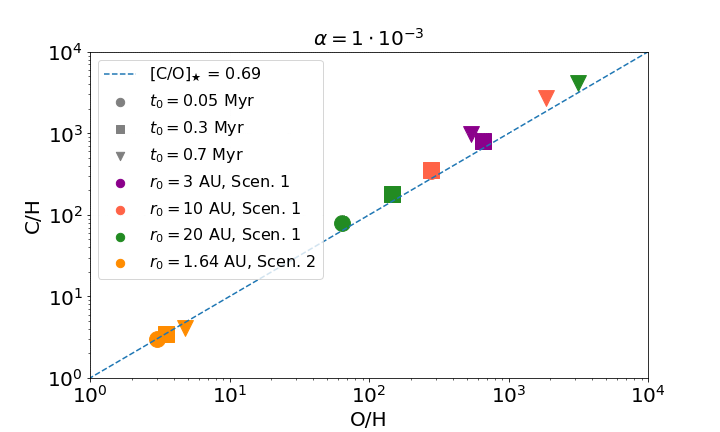}
    \caption{ C/H and O/H ratios of the  envelope of the inner planet are displayed for both scenarios, normalized to stellar composition. The blue diagonal line marks the stellar C/O ratio. A planet formed in scenario 1 shows slightly superstellar C/O content, while a planet formed in scenario 2 has significantly lower C/H and O/H ratios.}
    \label{fig:C/O ratio}
\end{figure}
We now investigate the water content of the inner planet within this formation scenario. In this scenario, the inner planet develops entirely within the water-ice line. For simplicity, migration of the inner planet was neglected in order to allow it to grow throughout the entire lifespan of the disk. For all simulations, the planet is positioned between the $\mathrm{Fe_3O_4}$ evaporation line and the water evaporation line, because $\mathrm{Fe_3O_4}$ is a crucial component for building protoplanets. The positions of these evaporation lines vary depending on the viscosity of the disk. Consequently, a different fixed position for the inner planet is chosen, corresponding to each viscosity. This results in fixed positions of 0.96 AU, 1.38 AU, and 1.64 AU (from low to high $\alpha$). Neglecting planet migration is an acceptable assumption when considering the water vapor content, because the water vapor of the  disk is radially evenly distributed within the water-ice line (see Fig. \ref{fig:water content disk + growthtracks scen 1}). Further discussion of this assumption can be found in Sect. \ref{discussion}.\\
In this scenario,  different insertion times between $t_0$ = 0.05 Myr and $t_0$ = 0.7 Myr are again tested. In Fig. \ref{fig:planet mass scen 3}, we illustrate the final mass of the inner planet for each simulation. Especially for $\alpha = 1 \cdot 10^{-4}$ and $\alpha = 5 \cdot 10^{-4}$, it is noticeable that the final planetary masses are lower compared to scenario 1. For $\alpha = 1 \cdot 10^{-3}$, the planets attain masses similar to HAT-P-11b, and a significant drop in planetary mass occurs only after an insertion time of 0.7 Myr. For lower $\alpha$-values, the observed mass of HAT-P-11b (indicated by the orange horizontal line in Fig. \ref{fig:planet mass scen 3}) is not reached for any formation time. This is attributed to the early formation of the gas giant (see Appendix \ref{appendix: gas giant}), which starts blocking pebbles early, leaving only a small amount of material in the inner disk (see Fig. \ref{fig:dust surface density}, bottom). Consequently, planetary growth is highly inefficient in the inner disk at low viscosities. However, for $\alpha = 10^{-3}$, the inner disk retains pebbles for a longer duration (Fig. \ref{fig:dust surface density}, bottom) due to the delayed formation of the giant planet, allowing efficient growth of early-forming planets. As the planetary mass is a more stringent constraint than the atmospheric abundances, we focus our discussion on simulations with $\alpha = 10^{-3}$ concerning planetary water content.\\
In Fig. \ref{fig:water content planet scen 3}, we present the final water content of the planetary envelope for different insertion times in a disk with $\alpha = 1 \cdot 10^{-3}$. The blue horizontal line represents the observed water content of HAT-P-11b within the range of uncertainty. The color code indicates the final planetary mass in Earth masses. The later the planet is inserted into the disk, the lower its final water content, which is due to the reduced water vapor content in the disk at later times (Fig. \ref{fig:water content disk + growthtracks scen 1}, bottom). Here, the observed water content of HAT-P-11b can be reproduced within the simulations if the planet is inserted after $t_0 \approx $ 0.6 Myr. However, the planet-insertion time is also limited by pebble supply. The planet must form before $t_0 \approx$ 0.8 Myr to accrete a sufficient mass. Overall, there is only a small time window for the planet to form late enough to grow in a disk where the water content has been sufficiently reduced while still having enough pebbles available to accrete sufficient mass.\\
Our simulations can thus reproduce the observed water content of HAT-P-11b within this scenario. The final mass of the simulated planet lies in a slightly higher range (between 30 and 50 $\mathrm{M_{\oplus}}$) than that indicated by observations of HAT-P-11b . Nevertheless, it seems possible to simultaneously replicate both the observed mass and water content by fine-tuning simulation parameters, such as envelope opacity (see Sect. \ref{sec:envelope opacity}).

\subsection{C/H and O/H ratio of the inner planet}
Not only does the water content of a planetary atmosphere provide constraints on its formation history, but carbon and oxygen abundances play a pivotal role in our understanding of the evolution of a planet and its system (e.g., \citealt{oberg2011effects, madhusudhan2017atmospheric, turrini2021tracing, bitsch2022drifting}). Therefore, we compare the C/O ratio of the planetary envelope in the two presented formation scenarios for $\alpha = 1 \cdot 10^{-3}$ in Fig. \ref{fig:C/O ratio}.\\
In general, a planet forming in the first scenario appears to have significantly higher C/H and O/H values compared to a planet developing in the second scenario. The C/O ratio for the first scenario results in values slightly above the stellar C/O ratio. For the second scenario, the C/H and O/H ratios are much lower, and the planetary C/O is similar to the stellar C/O ratio. The different insertion times seem to have little influence on the C/H and O/H values for scenario 2. However, for scenario 1, the planet's C/H and O/H values are more widely distributed, and are determined by its insertion time. Depending on the evolution time of the planet, different proportions of material have already drifted inward, allowing the planet to accrete pebbles and vapor with varying abundances.\\
On the other hand, including migration scenario 2 could affect the C/O ratio, as a migrating planet would cross several evaporation lines and therefore accrete different proportions of different compounds. For a more precise prediction of the C/O ratio of  HAT-P-11b, migration needs to be considered in future simulations. This subject is discussed in more detail in Sect. \ref{discussion} and also in Appendix \ref{appendix: C/O}.

\section{Discussion} \label{discussion}
Our simulations suggest that the most crucial factors determining the final planetary water content are the viscosity of the disk, the relative position of the gas giant to the water-ice line, and the timing of the 
growth of the inner planet. In this section, we discuss additional factors and how they could influence our results and interpretations.

\subsection{Disk temperature development} \label{discussion:temp}
The water vapor content of the inner disk depends on the relative position of the gap-opening planet with respect to the water-ice line, which in turn is influenced by the temperature of the disk. In our simulations, the time evolution of the temperature is neglected for simplicity \citep{schneider2021drifting}. In reality, disk temperature decreases over time because of the reduction in gas surface density and the declining luminosity of the central star \citep{pringle1981accretion, armitage2010astrophysics, bitsch2015structure}. Incorporating temperature evolution in the simulation would shift the evaporation lines over time toward the central star. This would add significant complexity to the simulations, as the shift of evaporation lines would effectively change the relative position of the gas giant to the water-ice line. However, as long as the gas giant does not cross the water evaporation line, the water content of the inner planet remains relatively unaffected, assuming the inner planet stays within the water-ice line at all times. In contrast, a migrating gas giant can cross multiple evaporation fronts during its growth (e.g., \citealt{madhusudhan2017atmospheric, schneider2021drifting}), altering the water content of the inner disk and consequently the inner planet. If indeed the gas giant crosses the water evaporation line, the inner planet could accumulate a high water content in its envelope because of the evaporation of water-ice-rich pebbles that were initially blocked by the gas giant.

\subsection{Disk mass and radius}
Our simulations adopt the same disk parameters regarding radius and mass as those of \cite{schneider2021drifting}, for simplicity. However, increasing the mass of the disk while keeping the radius constant would result in a denser disk with higher temperatures. Similarly, decreasing the total radius of the disk while maintaining the same mass would also create a denser disk with higher temperatures. Such temperature changes would, once again, lead to shifts in the evaporation lines (see \ref{discussion:temp}). \\
Additionally, the planetary accretion rates of pebbles and gas are influenced by the disk properties. The Stokes number is determined by the mass and radius of the  disk, which therefore influence the pebble-accretion radius of the planet \citep{morbidelli2015great, johansen2017forming}. For a fixed disk mass, a smaller disk radius would result in a faster depletion of the pebble reservoir because the formation of pebbles takes longer at greater orbital distances. This would increase the pebble surface density in the inner disk, thereby enhancing planet growth. Consequently, the outer planet would block the inward flow of pebbles earlier, leading to less enrichment of water vapor in the inner disk. In contrast, a larger disk radius would lead to a lower initial pebble flux because pebbles have to drift a greater distance to reach the inner boundary of the disk, resulting in slower growth.

\subsection{Disk vertical mixing}
The vertical mixing $\alpha_z$ describes the vertical range over which pebbles are distributed in the disk and is set here to $\mathrm{\alpha_z} = 1 \cdot 10^{-4}$ \citep{schneider2021drifting}. This vertical settling value is in agreement with observations \citep{dullemond2018disk,flaherty2018turbulence, pinilla2021growing}, but larger values for $\alpha_z$ cannot be excluded. A larger value for $\alpha_z$ would specifically result in a greater scale height for pebbles, $H_{peb}\propto \sqrt{\alpha_z}$, which describes the vertical distribution of pebbles, where lower scale heights mean that the pebble flux is more concentrated in the midplane of the disc. As an accreting planet is vertically centered in the disk, this would lead to the accretion of a lower fraction of passing pebbles, resulting in lower planetary masses. Moreover, as more pebbles were to pass the planet, the water content of the inner disk would increase even further, because a larger fraction of water-ice-rich pebbles would be able to move toward the inner disk and evaporate.

\subsection{Planetary envelope opacity} \label{sec:envelope opacity}
The gas-accretion rate of a planet is effectively determined by its envelope opacity $k_\mathrm{env}$ \citep{ikoma2000formation, machida2010gas, mordasini2015global, bitsch2021influence, brouwers2021planets, ndugu2021probing}. Higher $k_\mathrm{env}$ values result in a more opaque envelope, causing the planet to cool more slowly and therefore to accrete less gas. The envelope opacity determines how much gas the planet accretes and, consequently, whether or not it becomes a gas giant. In other words, $k_\mathrm{env}$ directly influences the final mass of the planetary envelope and therefore the total planet mass. We chose the envelope opacity so that, for all viscosities and formation times in a given scenario, the inner planet does not become a gas giant in the simulation, and the final mass matches the observations of HAT-P-11b as closely as possible.\\
Minor changes in the envelope opacity would slightly alter the envelope mass, while significant changes would have a more substantial impact. Specifically, very low $k_\mathrm{env}$ values would lead to the planet becoming a gas giant. Increasing $k_\mathrm{env}$ would mean that less gas were accreted overall, but the composition of the accreted gas would remain the same. This implies that the envelope opacity primarily influences the planet's mass but not the chemical composition of the envelope, as long as the planet remains in the same disk regions.\\
In scenario 1, the gas accretion rate influences the planet's water content. Here, the protoplanetary envelope already becomes enriched in water during pebble accretion and must be counteracted during gas accretion by accreting water-poor gas. However, before gas accretion begins, the water content of the  envelope ranges from $X_\mathrm{H_2O} = 10^2$ to $X_\mathrm{H_2O} = 10^3$ and cannot easily be compensated for by gas accretion (see Fig. \ref{fig:water content disk + growthtracks scen 1} and \ref{fig:water content planet scen 1}). To achieve a water-poor envelope for scenario 1, significantly more gas with substellar water content would need to be accreted at later stages, but this could only occur if the inner planet were to become a gas giant.\\
For the first scenario, the critical factor is the fraction of the accreted pebbles that contribute to the preliminary envelope. Within the simulations, it is assumed that 10\% of the accreted material during pebble accretion contributes to the planetary envelope. Detailed models by \cite{brouwers2020planets} and \cite{ormel2021planets} indicate that up to 50 \% of the pebbles could contribute to the envelope. This would make it even more challenging to achieve a substellar water content within scenario 1.\\
In scenario 2, the planet originates within the water-ice line and can therefore only accrete water vapor during gas accretion. Slight variations in envelope opacity and therefore different gas accretion rates are not expected to change the final water content in the planetary envelope because the abundance of the accreted gas remains the same. However, with a different gas-accretion rate, the final planetary mass can be fine-tuned to match the  mass of HAT-P-11b.\\
In conclusion, changing the envelope opacity could alter the final planet mass. By adjusting the envelope opacity in individual simulations, it appears possible to achieve the desired mass without significantly affecting the water content, which is particularly crucial for scenario 2. It is worth noting that we use the envelope opacity as a proxy to control the gas-accretion rate. In reality, gas accretion is a complex process that has been studied in 1D models (e.g., \citealt{mordasini2014grain, ormel2014atmospheric, marleau2017planetary, brouwers2021planets}) and 3D models (e.g., \citealt{klahr20063d, ayliffe2009gas, morbidelli2014meridional, cimerman2017hydrodynamics, lambrechts2017reduced, schulik2019global, moldenhauer2022recycling}), which consider detailed physics and dynamics around the planet, including recycling flows that might hinder efficient gas accretion onto Neptune-sized planets in inner disk regions (e.g., \citealt{cimerman2017hydrodynamics, lambrechts2017reduced, moldenhauer2022recycling}). This aligns with the slow gas-accretion rates used in our work.

\subsection{Recycling flows in the envelope}
Recycling flows could effectively reduce the water content in the planetary envelope (e.g., \citealt{cimerman2017hydrodynamics, lambrechts2017reduced}). A planet accreting large masses of pebbles is strongly heated in this process, as is its envelope. When the planet accretes water-ice-containing pebbles, these pebbles are heated when entering the planetary envelope, and consequently volatile  water-ice can evaporate in the outer layers of the envelope. These outer envelope layers can be transported away by so-called recycling streams and released back into the disk \citep{johansen2021pebble}, leading to a reduction in water content in the planetary envelope. This effect is not considered in the simulations and, depending on its strength, could decrease the water content of the planet in scenario 1.

\subsection{Equilibrium chemistry}
In this work, we assume that the water content of the atmosphere of  HAT-P-11b is equal to the amount of water accreted by the planet. However, equilibrium chemistry is expected to play a role in the planetary atmosphere and may alter its chemical composition. Equilibrium chemistry models suggest that refractory oxides can form in the atmosphere within this regime, leading to an oxygen deficit (e.g., \citealt{fonte2023oxygen}). Therefore, the actual water content of the atmosphere of  HAT-P-11b could be lower than the amount of water that it has accreted. To account for this, detailed modeling of atmospheric chemistry is necessary, which is beyond the scope of this study.\\
Furthermore, our model considers different chemical species that influence the C/O ratio of the  disk (see Appendix \ref{appendix: C/O}). The inward drift and evaporation of these species change the C/O ratio in the gas phase over time as carbon-rich material drifts inward. However, our model does not include chemical evolution of solids because the drift timescale is shorter than the chemical reaction timescale, preventing efficient alteration of the chemical composition of pebbles \citep{Booth2019, Eistrup2022}. The chemical evolution of the gas phase, on the other hand, might influence the composition of the gas in the inner regions. However, early carbon delivery, as in scenario 2, results in substellar C/O ratios within 1 Myr \citep{cridland2019connecting}, which is consistent with the C/O ratio evolution of our model (Fig.~\ref{fig:C/O ratio disk}).

\subsection{Contributions of solid accretion}
In reality, also planetesimals would form in protoplanetary disks (see \citealt{johansen2014multifaceted} for a review). Planetesimals in the outer disk regions containing water-ice could be scattered inward by a giant planet, which is similar to the water delivery process to Earth (e.g., \citealt{raymond2017origin}). However, the amount of water delivered to Earth is relatively small (e.g., \citealt{o2018delivery}). Despite HAT-P-11b having a considerably greater mass than Earth, its proximity to the star significantly reduces its capture radius, leading to a substantial decrease in the expected water delivery through planetesimal accretion from the outer system.\\
Another possibility would be the formation of a water-ice-rich outer planet that migrates inward and collides with an inner rocky planet (e.g., \citealt{raymond2018migration}), resulting in a planet with a moderate amount of water. However, such collisions might strip the planetary envelope (e.g., \citealt{liu2015giant, inamdar2016stealing, biersteker2019atmospheric}), resulting in a planet without a significant atmosphere. This is inconsistent with mass and radius measurements for  HAT-P-11b, which indicate the presence of an atmosphere (e.g., \citealt{zeng2019growth}).\\
\\

\section{Conclusion} \label{conclusion}
The water content of HAT-P-11b is reproduced within the simulations using scenario 2, which incorporates a gas giant in the disk capable of blocking water-ice-rich pebbles beyond the water evaporation line. Additionally, a high $\alpha$-value is necessary to facilitate the formation of a sufficiently massive inner planet. In the first scenario, where the sub-Neptune forms in the outer disk areas and migrates inward (without the presence of a gas giant), the mass of HAT-P-11b can be accurately reproduced, but the water content of the formed planets is too high compared to observations.\\
\\
To best reproduce the observational data from HAT-P-11b, the following conditions should be applied:
\begin{itemize}
\item A gas giant beyond the water-ice line is required to effectively reduce the water content in the inner disk by blocking the inwardly drifting ice-bearing pebbles (scenario 2).
\item Fast gas diffusion is essential for the rapid removal of water-rich gas inside the disk, which is achieved with a high $\alpha$-viscosity for the disk.
\item The gas giant should begin to form early, ensuring an early reduction in water vapor content in the inner disk. In this study, the gas giant begins to grow simultaneously with the beginning of the evolution of the protoplanetary disk.
\item The inner planet should form early in order to accrete sufficient material (insertion times $t_0 <$ 0.8 Myr). Simultaneously, the inner planet should begin accreting after $t_0 \approx$ 0.6 Myr, when the disk is sufficiently depleted in water vapor.
\end{itemize} 

\noindent The observed mass and water abundances of HAT-P-11b suggest that the outer gas giant efficiently blocked water-ice from reaching the inner system. The efficiency of this process depends strongly on the formation time of the gas giant, with our simulations indicating early giant planet formation. This aligns with the idea that Jupiter, in our Solar System, formed early and thus separated carbonaceous and non-carbonaceous materials (e.g., \citealt{kruijer2017age, morbidelli2020subsolar}). 
Moreover, the significantly greater mass of HAT-P-11b  compared to terrestrial planets in the Solar System implies that more material was required to enter the inner HAT-P-11 system compared to the solar system (e.g., \citealt{izidoro2021effect}). This suggests that HAT-P-11c formed later than Jupiter to allow sufficient material to pass. Alternatively, the embryos of the terrestrial planets in the Solar System could have formed at later times when the pebble flux in the inner system was much lower.\\
High viscosity is favored because it allows faster gas diffusion and water vapor content is reduced earlier in the disk. At high viscosity, pebbles are smaller, resulting in slower drift, enabling the planet to accrete them over a longer period of time, thus achieving higher planetary masses. By fine-tuning individual parameters, it seems possible to simultaneously reproduce both the observed mass and water content of HAT-P-11b.\\
While our simulations provide valuable insights into constraining the formation history of the HAT-P-11 system, additional information could be used to further constrain the formation processes. Considering that not only the water content of the planetary envelope differs among different formation scenarios, but also the C/H and O/H ratios, future observational data for the composition of the envelope of HAT-P-11b  will be crucial in further refining our understanding of the formation history of the system, making HAT-P-11b a prime target for JWST observations.\\ 
\\

\noindent
\small{\textit{Acknowledgements.} B.B. thanks the European Research Council (ERC Starting Grant 757448-PAMDORA) for their financial support. A.D.S. acknowledges funding from the European Union H2020-MSCA-ITN-2019 under Grant no. 860470 (CHAMELEON). This research has made use of data obtained from or tools provided by the portal of The Extrasolar Planets Encyclopaedia (\footnotesize \url{http://www.exoplanet.eu}).}

\bibliographystyle{aa}
\bibliography{literature}

\appendix
\section{Molecular abundances}
In Table~\ref{tab:chem model}, we display the distribution of the different elements into solids within our model.
\begin{table*}
   \centering
    \begin{tabular*}{0.95\textwidth}{@{\extracolsep{\fill}}ccc}
    \hline
    \hline
    Species & Evap.temp. $T_{evap}$ [K] & volume mixing ratio\\
    \hline
    $\mathrm{CO}$ & 20 & 0.2 $\times$ C/H\\
    $\mathrm{N_{2}}$ & 20 & 0.45 $\times$ N/H\\
    $\mathrm{CH_4}$ & 30 & 0.1 $\times$ C/H\\
    $\mathrm{CO_2}$ & 70 & 0.1 $\times$ C/H\\
    $\mathrm{NH_3}$ & 90 & 0.1 N/H\\
    $\mathrm{H_2S}$ & 150 & 0.1 $\times$ S/H\\
    \makecell{$\mathrm{H_2O}$\\ \\ } & \makecell{150 \\ \\} & \makecell{\\O/H - (3 $\times$ $\mathrm{MgSiO_3}$/H + 4 $\times$ $\mathrm{MgSiO_4}$/H + CO/H + 2 $\times$ $\mathrm{CO_2}$/H \\
    + 3 $\times$ $\mathrm{Fe_2O_3}$/H + 4 $\times$ $\mathrm{Fe_3O_4}$/H + VO/H + TiO/H + 3 $\times$ $\mathrm{Al_2O_3}$ \\
    + 8 $\times$ $\mathrm{NaAlSi_3O_8}$
    + 8 $\times$ $\mathrm{KAlSi_3O_8}$)}\\ \\
    $\mathrm{Fe_3O_4}$ & 371 & $\frac{1}{6}$ $\times$ (Fe/H - 0.9 $\times$ S/H)\\
    C (carbon grains) &  631 & 0.6 $\times$ C/H\\
    FeS & 704 & 0.9 $\times$ S/H\\
    $\mathrm{NaAlSi_3O_8} $ & 958 & Na/H\\
    $\mathrm{KAlSi_3O_8} $ & 1006 & K/H\\
    $\mathrm{Mg_2SiO_4} $ & 1354 & Mg/H - (Si/H - 3 $\times$ K/H - 3 $\times$ Na/H)\\
    $\mathrm{Fe_2O_3} $ & 1357 & 0.25 $\times$ (Fe/H - 0.9 $\times$ S/H)\\
    VO & 1423 & V/H\\
    $\mathrm{MgSiO_3}$ & 1500 & Mg/H - 2 $\times$ (Mg/H - (Si/H - 3 $\times$ K/H - 3 $\times$ Na/H))\\
    $\mathrm{Al_2O_3} $ & 1653 & Al/H - (K/H + Na/H)\\
    TiO & 2000 & Ti/H\\
    \hline
    \hline
    \multicolumn{3}{c}{}
    \end{tabular*}
    \caption{The chemistry model (\citealt{bitsch2020influence, schneider2021drifting2}) provides the distribution of elements among molecules. Here, it is assumed that carbon grains are formed from 60 \% of the available carbon.}
    \label{tab:chem model}
\end{table*}

\section{Implementation of a gas giant} \label{appendix: gas giant}

\begin{figure}[h]
    \centering
    \includegraphics[width=0.5\textwidth]{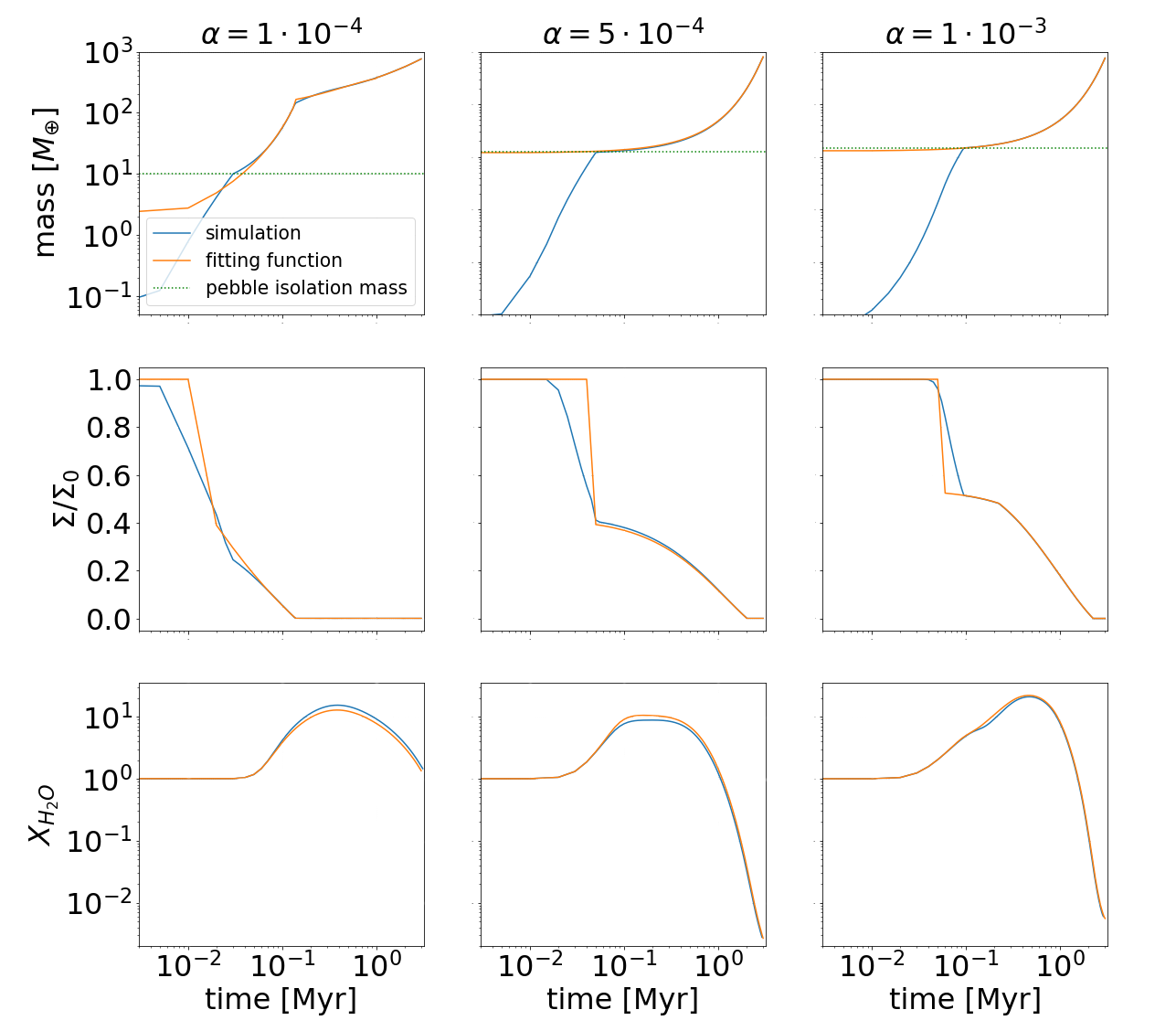}
    \caption{Comparison of simulation results for a fully simulated gas giant (blue) and simulations using the fitted mass function (orange) presented for different $\alpha$-values. The top panel shows the mass of the gas giant over time, with the green vertical line denoting the pebble isolation mass. The center panel displays the reduced disk surface density due to the gas giant. The bottom panel shows the water vapor content of the disk  inside water-ice line at 0.3 AU.}
    \label{fig:gas giant details}
\end{figure}

To simulate the effect of the mass of the  gas giant on the disk, a planet was simulated for every viscosity. The gas giant was inserted at $t_0$ = 0.0 Myr, right at the beginning of the lifetime of the disk. The envelope opacity was adjusted so that that the gas giant reaches roughly the observed mass of  HAT-P-11c, of namely 731 $\mathrm{M_{\oplus}}$, by the end of thes lifetime of the disk, which is 3 Myr.\\
In Fig. \ref{fig:gas giant details}, we present a comparison of the fully simulated gas giant (blue) and the simulation results using the determined fitting function (orange) for different $\alpha$-values. The upper row displays the planet mass growth function over time, with the green horizontal line marking the pebble isolation mass. Our primary focus is on the effect the gas giant has on the disk once it reaches the pebble isolation mass, making the fit accuracy relevant at this stage within this work. In the middle row, we depict the reduced disk surface density at the position of the  planet due to gap opening, normalized to an unperturbed disk. For instance, $\Sigma$/$\Sigma_0$ = 0.5 corresponds to a surface density reduced by half of the initial surface density. A gap is considered sufficiently deep enough to block inward drifting pebbles when $\Sigma$/$\Sigma_0$ $\approx 0.5$. The bottom row illustrates the resulting water content of the disk inside the water-ice line at 0.3 AU, clearly demonstrating that our fitting function produces comparable water contents when compared to a real growing planet.

\section{Scenario 3: Gas giant inside the water evaporation line} \label{appendix: scen3}
\begin{figure}[h]
    \centering
    \includegraphics[width=0.51\textwidth]{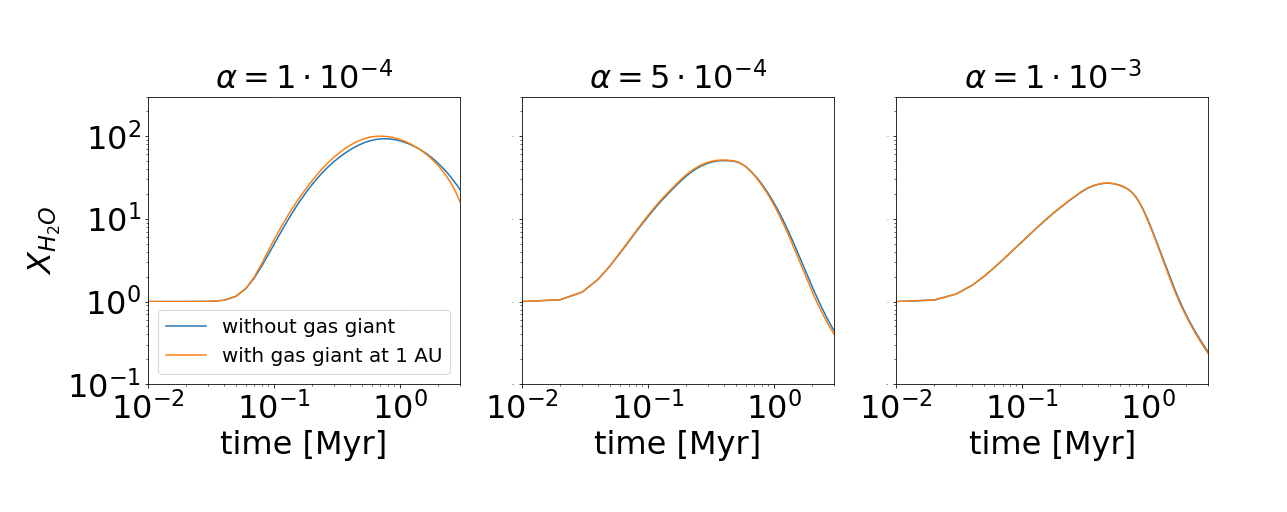}
    \caption{Water vapor content in the disk for different $\alpha$-values without a gas giant (blue) and with a gas giant inside the water-ice line (orange) according to scenario 3. The water vapor content in the disk does not differ between these two scenarios. Hence, the gas giant located inside the water-ice line does not influence the resulting water vapor content in the disk.}
    \label{fig:scen3 disk}
\end{figure}

For this additional scenario, the gas giant is located within the water-ice line, and therefore the position at 1 AU was chosen as it is inside the evaporation line for all $\alpha$-values tested. In Fig. \ref{fig:scen3 disk}, we provide a comparison of the water vapor content in the disk inside the water-ice line for a scenario without a gas giant (blue) and with a gas giant located inside the water-ice line (orange). The presence of the gas giant does not influence the water vapor content in the disk inside the water-ice line compared to scenario 1. This is because pebbles still cross the water-ice line and consequently evaporate. Therefore, a planet evolving in this scenario accretes water-rich gas after all, and thus, this scenario cannot account for the water-poor atmosphere of HAT-P-11b.

\section{C/O ratio of the gas in the disk} \label{appendix: C/O}
\begin{figure*}[h]
    \centering
    \includegraphics[width=0.98\textwidth]{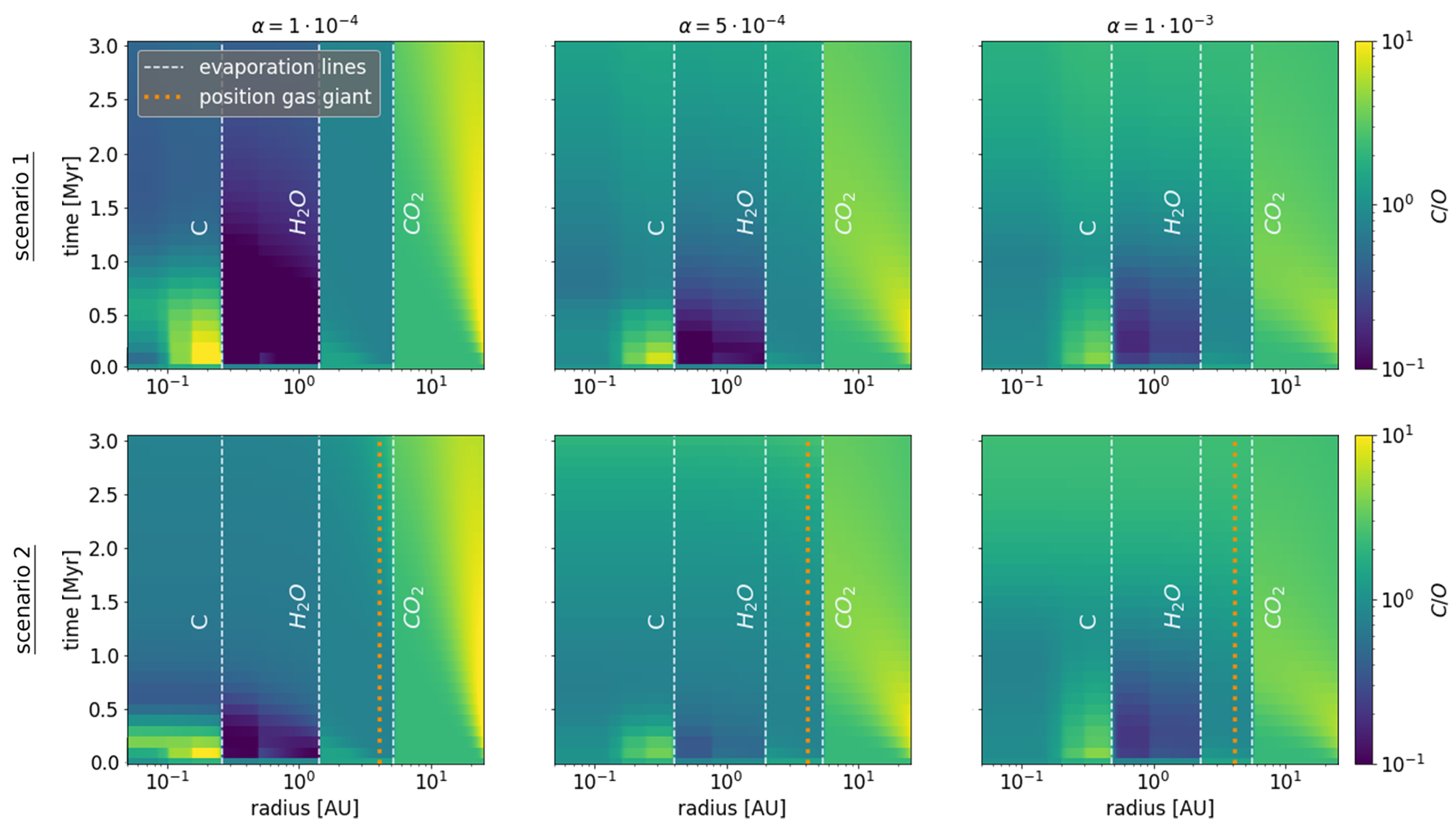}
    \caption{Disk C/O ratio in the gas phase normalized to stellar values for different $\alpha$-viscosities for scenario 1 (top) and scenario 2 (bottom), as functions of disk radius and time. Different evaporation lines are marked in white. The orange vertical line indicates the position of the gas giant for scenario 2.}
    \label{fig:C/O ratio disk}
\end{figure*}

The gas giant not only influences the water vapor content in the disk but also the abundances of other species, depending on its relative position to certain evaporation lines. Figure \ref{fig:C/O ratio disk} shows the C/O ratio of the gas normalized to stellar values in disks with different $\alpha$-viscosities for scenario 1 (top) and scenario 2 (bottom) as a function of disk radius and time. The formation of the outer gas giant in scenario 2 affects the disk's C/O ratio within its orbit. In the scenario without the outer giant, the initial C/O ratio interior to the water-ice line is much lower (top in Fig. \ref{fig:C/O ratio disk}) due to the evaporation of water-ice-rich pebbles, compared to the scenario where the pebbles are blocked by the gas giant. In contrast, carbon-rich gas diffusing into the inner disk compensates for the low C/O ratio caused by evaporated water-ice. Conversely, the outer giant also blocks carbon-rich grains, resulting in a slightly lower C/O ratio in the very inner disk compared to the situation without an outer planet. These trends appear to be independent of the viscosity of the disk, with more extreme C/O ratios reached at lower viscosities.

\end{document}